\documentclass[onecolumn,amsmath,amssymb,aps,pra]{revtex4-2}

\usepackage{graphicx}
\usepackage{dcolumn}
\usepackage{bm}
\usepackage{physics}
\usepackage{hyperref}
\usepackage{placeins}
\begin{document}
	
	\title{Quantum-RAM Implementation Using Multiple Interacting Rydberg-Blockaded EIT Systems}
	
	\author{Avirup Chakraborty}
	\affiliation{S.N. Bose National Centre for Basic Sciences, \\JD Block, Sector-III, Bidhannagar, Kolkata, West Bengal-700106, India.}
	\email{avirup.chakraborty@bose.res.in}  
	
	\author{Shrabana Chakrabarti}
	\affiliation{Sister Nivedita University, \\DG 1/2, DG Block(Newtown), Action Area I, Newtown, Kolkata, Chakpachuria, West Bengal 700156, India.}
	\email{shrabana.cs@snuniv.ac.in}
	
	\date{\today}
	
	\begin{abstract}
		We propose a novel theoretical architecture for implementing a quantum random access memory (qRAM) based on quantum random walks in a Rydberg blockaded atomic ensemble utilizing multilevel Electromagnetically Induced Transparency (EIT). Unlike previous approaches that rely on geometric phase gates in solid-state or trapped ion systems, our scheme harnesses the strong, coherent dipole–dipole interactions between Rydberg atoms to achieve high-fidelity phase control of photonic qubits without the need for cryogenic temperatures. By generating conditional phase shifts through cross-phase modulation in multiple lambda-type EIT systems, we realize the controlled unitary operations requisite for an efficient qRAM. In the proposed architecture, Zeeman splitting is used to engineer a set of parallel lambda systems in a cavity, where pairs of magnetic sublevels of the ground state are coupled to highly excited Rydberg states via circularly polarized laser pulses. These Rydberg excited EIT systems serve as the elementary phase gates that form the nodes of a binary tree enabling quantum random walking. Address and data qubits are encoded into distinct probe fields and coherently mapped into the metastable atomic states, where their interactions within the EIT medium generate conditional phases required for state-selective routing. The system uses $n+m$ layers of cold alkali atoms to form an $n$-level binary tree of Rydberg nodes connected to $2^n$ cavity-trapped memory atoms, operated by $n+m$ laser pulses acting as quantum walkers and address units. Our scheme offers a scalable, reducing operational complexity to $\mathcal{O}(n)$ and highly coherent pathway toward photonic qRAM, exploiting collective Rydberg interactions to realize programmable, parallel entangling operations in an atomic ensemble.
	\end{abstract}
	
	
	\maketitle
	
	\section{Introduction}
	
	The foundations of quantum computation, initiated by Turing machines \cite{benioff1980}, were advanced by Feynman's simulations \cite{feynman1982} and Deutsch's parallelism \cite{deutsch1985}. Subsequent demonstrations of computational superiority \cite{bernstein1993, berthiaume1992}, solidified by Shor's and Grover's algorithms \cite{shor1994, grover1996}, profoundly impacted artificial intelligence and machine learning. Quantum algorithms yield exponential speedups; for example, the HHL algorithm reduces vector clustering complexity from classical $\mathcal{O}(\text{poly}(MN))$ to $\mathcal{O}(\log(MN))$ \cite{harrow2009, lloyd2013, aaronson2015}. However, practical implementation suffers from severe data loading overhead \cite{lloyd2013, biamonte2017}.
	
	To address this, Giovannetti proposed quantum random access memory (qRAM) enabling coherent superposition access \cite{giovannetti2008_1, giovannetti2008_2, giovannetti2008_3}. Given the address superposition $\sum_j \alpha_j |j\rangle_a$, the qRAM returns the correlated data $\sum_j \alpha_j |j\rangle_a |D_j\rangle_d$. Giovannetti introduced two architectures: fanout and routing buses along $2^n$ paths via controlled unitaries, and bucket brigades, utilizing distributed qutrits updated in $n=\log(N)$ steps. Implementations in optical lattices and cavity arrays remain challenging \cite{duan2003, farhi1998}, whereas optical fanout uses atomic states, cavity bucket brigades utilize Raman-updated qutrits \cite{giovannetti2008_2}. Crucially, fanout requires exponential gates, exacerbating decoherence, whereas bucket brigades require $\mathcal{O}(n)$ qutrit coherence with $\mathcal{O}(n^{-2})$ error tolerances.
	
	To resolve these limitations, quantum random walks—unitary Markov chain analogues—are proposed as a foundational architecture \cite{aharonov1993, ambainis2001, aharonov2001, childs2009}. They provide algorithmic universality and replace fragile node gates with robust chirality-driven walkers, reducing complexity to $\mathcal{O}(n)$ steps. Extending Asaka et al.'s discrete-time binary tree walk \cite{asaka2021, asaka2023_1, asaka2023_2}, the scheme employs four Hilbert spaces: address $V_A$, bus $V_B$, chirality $V_C$, and data $V_D$. Internal bucket states $|L\rangle$ and $|R\rangle$ direct routing across $2^n$ nodes. For address superposition $|a_{n-1} \dots a_0\rangle_A$ and data $|x_{m-1}(a) \dots x_0(a)\rangle_D$, the walk operates in $V = V_B \otimes V_C \otimes V_A \otimes V_D$ in three sequential stages:
	\begin{equation}
		\sum_{a \in A} |0,0\rangle_B |0\rangle_C |a\rangle_A |0\rangle_D \mapsto \sum_{a \in A} |0,0\rangle_B |0\rangle_C |a\rangle_A |x(a)\rangle_D
	\end{equation}
	
	Given a binary tree of depth $n$ with $2^n$ nodes, the protocol executes sequentially. First, routing: for address $|a_{n-1}, \dots, a_0\rangle_A$, information is stored in chirality state $|0\rangle_C$. A shift operator acts on bus vector $|w,l\rangle_B$ at depth $l$, moving it to child node $|2w,l+1\rangle_B$ if the internal state is $|0\rangle_C$, or $|2w+1,l+1\rangle_B$ if $|1\rangle_C$. Consequently, traversing from root $|0,0\rangle_B$ requires applying unitary operators across tensor state $|0,0\rangle_B |0\rangle_C |a\rangle_A |0\rangle_D$. Second, querying: reaching memory addresses entangles the bus with data state $|x(a)\rangle_D$ corresponding to $|a\rangle_A$. Third, output: due to overall unitarity, reversing operations from $|a,n\rangle_B |a_0\rangle_C |a\rangle_A |x(a)\rangle_D$ successfully retrieves the queried data.
	
	This architecture eliminates active gates and the need for coherence across $n \times 2^n$ qutrits, reducing the data query complexity to $\mathcal{O}(n)$ using a four-channel qubit state \cite{asaka2021, asaka2023_1, asaka2023_2}.
	
	Realizing quantum walks without geometric phase gates requires robust unitary transformations. Early implementations in calcium ion traps \cite{schmidt2003}, beryllium traps \cite{leibfried2003}, nuclear spins \cite{kane1998}, and Josephson junctions \cite{pashkin2003} faced cryogenic constraints and decoherence. Photonic platforms offer superior environmental isolation \cite{obrien2003, petrosyan2005}. While EIT dark states, polarization CNOT gates \cite{andre2002, gasparoni2004}, and optical lattice quantum walks \cite{agarwal2005, cote2006, jaksch2000} show promise, photon interactions remain weak. Employing Rydberg dipole-dipole couplings \cite{gallaghar1994} enables high-fidelity entangled states for scalable qRAM \cite{moller2008}, while periodic EIT modulation facilitates cross-phase shifts \cite{friedler2005} and robust conditional phase gates \cite{shahmoon2011, petrosyan2012, paredes2014}.
	
	Similar to the approach adopted by Asaka et al. \cite{asaka2023_2}, this optimized ensemble comprises $n+m$ cold alkali atom layers forming an $n$-level binary tree connected by hollow-core waveguides, interfacing with the $2^n$ cavity memory atoms via probe and signal pulses, restricting operational complexity to $\mathcal{O}(n)$.
	
	\section{Methodology}
	
	\subsection{Generation of Conditional Phase in Quantum Walkers using D-D Interaction}
	The $n$ address state $|a\rangle_A = |a_{n-1}, \dots, a_0\rangle_A = |a_{n-1}\rangle_{A_{n-1}} \otimes \dots \otimes |a_0\rangle_{A_0} (a_i \in \{0,1\}; 0 \leqslant i \leqslant n - 1)$ is stored as the polarization states of the $n$ address walker probe pulses, conventionally representing left and right circularly polarized photons as $|0\rangle$ and $|1\rangle$, respectively ($|L\rangle \to |0\rangle; |R\rangle \to |1\rangle$). The $m$ data walker probe pulses comprise a tensor product of Fock states $|0\rangle_D \to |0\rangle_{A_n} \otimes \dots \otimes |0\rangle_{A_{n+m}}$, utilized for retrieval of data states $\left|x^{(a)}\right\rangle_D = \left|x^{(a)}_{m-1}, \dots, x^{(a)}_0\right\rangle_D = \left|x^{(a)}_{m-1}\right\rangle_{D_{m-1}} \otimes \dots \otimes \left|x^{(a)}_0\right\rangle_{D_0} \left(x^{(a)}_i \in \{0,1\}; 0 \leqslant i \leqslant m - 1\right)$ stored as ground-state superpositions $|\psi\rangle = \alpha|g\rangle + \beta|e\rangle$ of three-level atoms in $2^n$ memory locations.
	
\begin{figure}[htbp]
		\centering
		\includegraphics[width=0.45\linewidth]{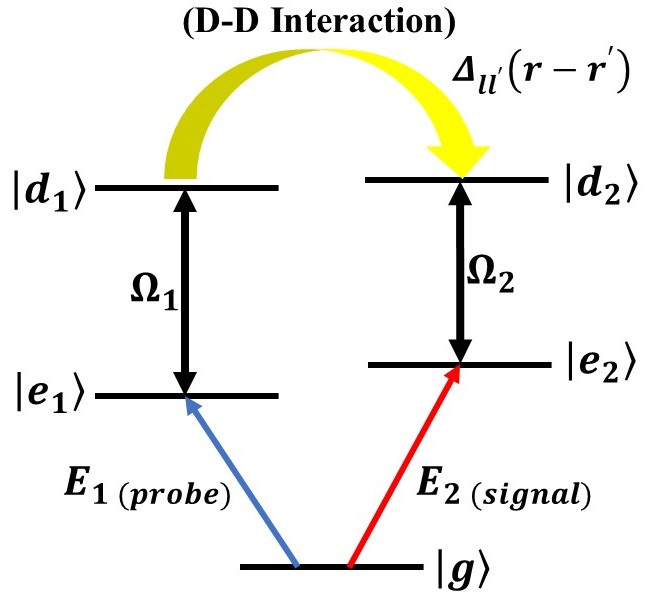}
		\caption{EIT Dipole-Dipole Interaction Dressed State Diagram.}
		\label{fig:fig1}
	\end{figure}

	A ladder-type electromagnetically induced transparency (EIT) system (See Fig.~\ref{fig:fig1}) involving excited Rydberg states, introduced by Friedler et al., enables phase-dependent routing \cite{friedler2005_2}. The architecture comprises a cylindrically symmetric hollow-core waveguide of length $L$ filled with $N$ cold atoms initially prepared in the ground state $|g\rangle$. Trapped atoms possess a Gaussian density $\rho(\mathbf{r}) = (\pi w_a^2)^{-1} e^{-r_\perp^2/w_a^2}(N/L)$ with a width $w_a \leq w_f$. Two counterpropagating weak quantum fields $\hat{E}_{1,2}$ propagate along the $z$-axis with a transverse Gaussian width $w_f$, expressed as $\hat{E}_l(\mathbf{r}) = \epsilon_l e^{-r_\perp^2/2w_l^2} \hat{\mathcal{E}}_l(z)$ using the traveling-wave operator $\hat{\mathcal{E}}_l(z)$, radial distance $r_\perp$, and per-photon electric field $\epsilon_l = \sqrt{\hbar\omega_l / 2\epsilon_0 V}$ within quantization volume $V = \pi w_f^2 L$. Fields resonantly drive $|g\rangle \to |e_{1,2}\rangle$ with couplings $g_l = (\wp_{ge_l}/\hbar)\epsilon_l$ and transition dipole matrix elements $\wp_{ge_l}$. Two strong classical driving fields with Rabi frequencies $\Omega_{1,2}$ couple intermediate states $|e_{1,2}\rangle$ to highly excited states $|d_{1,2}\rangle$. Static field $E_{\text{st}}\mathbf{e}_z$ induces permanent dipoles $p = \frac{3}{2} n q e a_0 \mathbf{e}_z$, where $n, q$ denote quantum numbers, $e$ charge, and $a_0$ Bohr radius. Excited atoms at $r, r'$ interact via potential $V_{dd}$, creating a shift $\hbar\Delta_{ll'}(\mathbf{r} - \mathbf{r}') = C_{ll'} \frac{1-3\cos^2\vartheta}{|\mathbf{r}-\mathbf{r}'|^3}$, where $\vartheta$ is the angle between polarization $\mathbf{e}_z$ and relative vector $\mathbf{r}' - \mathbf{r}$. Parameter $C_{ll'} = \wp_{d_l}\wp_{d_{l'}}/(4\pi\epsilon_0)$ uses dipole moments $\wp_{d_l} = \langle d_l | p | d_l \rangle$ \cite{shahmoon2011, petrosyan2012}. Averaging over the volume $\Delta V$ containing $N_r = \rho(\mathbf{r})\Delta V \gg 1$ atoms defines the collective transition operators $\hat{\sigma}_{\mu\nu}(\mathbf{r}) = \frac{1}{N_r}\sum_{j=1}^{N_r} |\mu\rangle_j \langle\nu|$. In the frame rotating with optical field frequencies, the total interaction Hamiltonian governing the system is $H = H_{af} + H_{dd}$, encompassing both atom-field and dipole-dipole interaction Hamiltonian terms, defined as:
	\begin{align}
		H_{af} &= -\hbar \int d^3r \rho(\mathbf{r}) \sum_{l=1,2} \left[ g_l e^{-r_\perp^2 / 2w_f^2} \hat{\mathcal{E}}_l(z) \hat{\sigma}_{e_l g}(\mathbf{r}) + \Omega_l \hat{\sigma}_{d_l e_l}(\mathbf{r}) \right] + \text{H.c.} \tag{1a} \\
		H_{dd} &= \hbar \int d^3r \rho(\mathbf{r}) \int d^3r' \rho(\mathbf{r}') \times \frac{1}{2} \sum_{l,l'=1,2} \hat{\sigma}_{d_l d_l}(\mathbf{r}) \Delta_{ll'}(\mathbf{r}-\mathbf{r}') \hat{\sigma}_{d_{l'} d_{l'}}(\mathbf{r}') \tag{1b}
	\end{align}
	
	From the Hamiltonian $H$, we derive the Heisenberg-Langevin equation for the atomic operators $\hat{\sigma}_{e_l g}(\mathbf{r})$ and $\hat{\sigma}_{d_l e_l}(\mathbf{r})$, alongside the propagation equation for slowly varying signal and walker probe quantum fields $\hat{\mathcal{E}}_{1,2}$. Solving perturbatively in the small parameters $g_{1,2}\hat{\mathcal{E}}_{1,2}/\Omega_{1,2}$ within the adiabatic approximation yields the dark-state polariton propagation equation \cite{shahmoon2011, petrosyan2012}:
	\begin{equation}
		\left( \frac{\partial}{\partial t} \pm v_{1,2} \frac{\partial}{\partial z} \right) \hat{\Psi}_{1,2}(z,t) = -i \sin^2\theta_{1,2} \hat{S}_{1,2}(z,t) \hat{\Psi}_{1,2}(z,t) \tag{2}
	\end{equation}
	
\begin{figure}[htbp]
		\centering
		\includegraphics[width=0.7\linewidth]{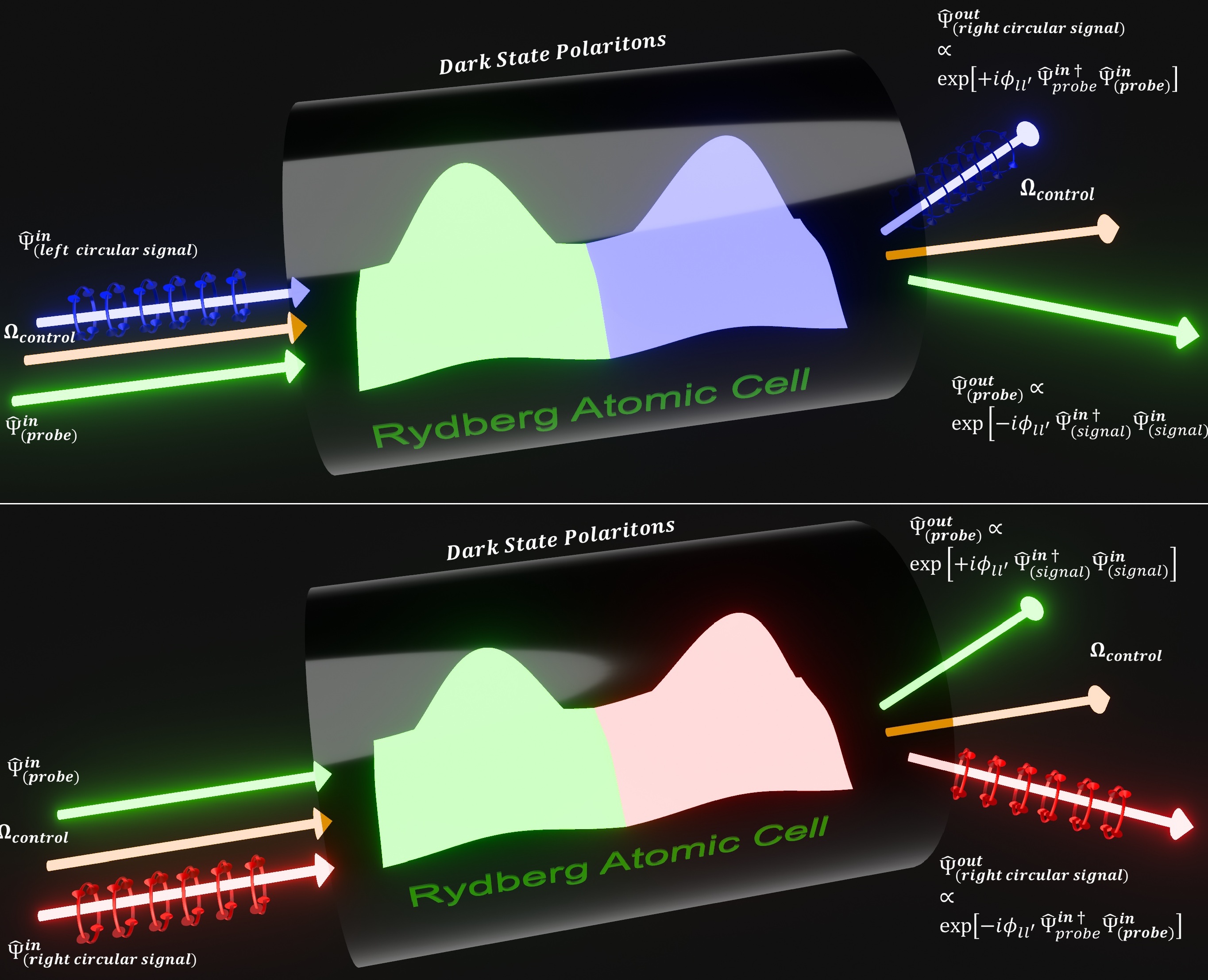}
		\caption{Dipole-Dipole Polariton Interaction inside Rydberg atomic cell and subsequent phase generation. The left (right) polarization state of signal pulses have been shown in (a) blue (top) and in (b) red (bottom).}
		\label{fig:fig2}
	\end{figure}

	Here, the polariton fields are $\hat{\Psi}_{1,2} = \sqrt{c/v_{1,2}} \hat{\mathcal{E}}_{1,2}$, mixing angles are $\tan^2\theta_l = \left(\frac{g_l^2 N}{|\Omega_l|^2}\right) \left(\frac{w}{w_a}\right)^2$ for $(l = 1,2)$, and effective width is $w = w_a w_f (w_a^2 + w_f^2)^{-1/2}$, where $w_a, w_f$ represent Gaussian distributions of the atomic ensemble and walker probe. The weak fields acquire cross-phase modulation:
	\begin{equation*}
		\hat{S}_l(z,t) = \frac{1}{L} \int_0^L dz' [ \Delta_{ll}(z - z') \sin^2 \theta_l \hat{I}_l(z', t) + \Delta_{ll'}(z - z') \sin^2 \theta_l \hat{I}_{l'}(z', t) ]
	\end{equation*}
	
	with excitation number operator $\hat{I}_l \equiv \hat{\Psi}_l^\dagger \hat{\Psi}_l = (c/v_l) \hat{\mathcal{E}}_l^\dagger \hat{\mathcal{E}}_l$, where $\Delta_{ll}$ and $\Delta_{ll'}$ denote 1D Rydberg dipole-dipole self- and neighbor-interaction potentials. The field solutions \cite{shahmoon2011, petrosyan2012} are
	\begin{equation}
		\hat{\Psi}_l(z,t) = \exp \left[ -i \sin^2 \theta_l \int_0^t dt' \hat{S}_l(z \mp v_l(t-t'), t') \right] \hat{\Psi}_l(z \mp v_l t, 0) \tag{3}
	\end{equation}
	Here, weak quantum fields acquire a time-dependent phase term through mutual signal and walker probe pulse interactions within transit time $t_{\text{out}} = L/v_l$, limited by atomic spin excitation operator $\hat{\sigma}_{d_l g}$ relaxation. This cross-phase modulation critically enables quantum routing. Because the cross-phase modulation term $\hat{S}_l(z,t)$ depends on the number operator $\hat{I}_{l'}(z', t)$, we write
	\begin{equation}
		\hat{S}_l(z_1, z_2, t) \approx -\sin^4 \theta \int_0^t dt' [ \Delta_{ll'}(z - z') \sin^2 \theta_l \hat{I}_{l'}(z', t) ] \tag{4}
	\end{equation}
	
	only if the dipole moment term for the address probe walker polariton is much smaller than that of the signal probe walker polariton ($C_{ll} < C_{ll'}$). If the polariton fields are written in terms of raising and lowering operators $\hat{\Psi}_{l,l'}(z,t)^\dagger$ and $\hat{\Psi}_{l,l'}(z,t)$, then the evolution of the polariton fields inside the waveguide medium can be represented by the coherent states $|\psi_{l,l'}\rangle \equiv \prod_k |\psi_{l,l'}^k\rangle$. The coherent states are taken as eigenstates of the polariton field operators such that $\hat{\Psi}_{l,l'}(z)|\psi_{l,l'}\rangle = \psi_{l,l'}(z)|\psi_{l,l'}\rangle$.
	\begin{equation}
		\langle \hat{\Psi}_l(L, L/v_l) \rangle = \psi_l(0) \langle \psi_{l'} | \exp \left[ -i \frac{\sin^4 \theta_l}{L} \int_0^{L/v_l} dt' \int_0^L dz' \Delta_{ll'}(z' - v_l t) \hat{I}_{l'}(z' + v_l t, 0) \right] | \psi_{l'} \rangle \tag{5}
	\end{equation}
	Doing the inner product: $\langle \psi_{l'} | \hat{I}_{l'}(z') | \psi_{l'} \rangle = \langle \psi_{l'} | \hat{\Psi}_{l'}^\dagger \hat{\Psi}_{l'} | \psi_{l'} \rangle = |\psi_{l'}|^2$ and substituting the value the integral $\frac{\sin^4 \theta_l}{L} \int_0^{L/v_l} dt' \int_0^L dz' \Delta_{ll'}(z' - v_l t) = \frac{C_{ll'} \sin^4 \theta}{\hbar w^2 v_l}$, calculated in \cite{shahmoon2011, petrosyan2012}, we obtain the expectation value of the outgoing address probe walker as:
	\begin{equation}
		\langle \hat{\Psi}_l(L, L/v_l) \rangle = \psi_l(0) \exp[i \phi_{ll'} \tilde{n}_{l'}]; \quad \phi_{ll'} = \frac{C_{ll'} \sin^4 \theta}{\hbar w^2 v_l}, \tilde{n}_{l'} = |\psi_{l'}|^2 \tag{6}
	\end{equation}
	The address walker probe field acquires a phase proportional to the average photon number $\tilde{n}_s$ of the signal probe coherent states $\Psi_s$. This phase generation due to cross-interaction between address and signal probe walkers makes selective routing possible based on the internal signal state (See Fig.~\ref{fig:fig2}). For our purpose, if we selectively set the average photon number of the signal polariton field based on its polarization state by using an appropriate polarizer, say: if we set two different excitation numbers $\hat{I}_{L,R}(z', t)$ for the signal polariton with left and right circular polarization states ($|L, R\rangle$), such that the accumulated phase of the address walker probe polariton is $\phi_{ll'} \tilde{n}_{l'} \approx \frac{\pi}{2}, \frac{3\pi}{2}$, such that we obtain $\langle \hat{\Psi}_l(L, L/v_l) \rangle = \psi_l(0) \exp\left[i \frac{\pi}{2}\right] \left(\exp\left[i \frac{3\pi}{2}\right]\right) = +i\psi_l(0)(-i\psi_l(0))$. By differentially choosing average photon numbers $\tilde{n}_{l;L} = 1$ and $\tilde{n}_{l;R} = 3$ for left and right circularly polarized signal states $|\psi_v\rangle = |L, R\rangle$, the address walker acquires equal and opposite phase states $|\psi_l\rangle = |+, -\rangle = \pm i\psi_l(0)$. Following the interaction time, applying an appropriate control laser retrieves the walker probe along positive or negative waveguide directions with velocities $\pm v$ and phases $\pm \frac{\pi}{2}$. This process constitutes a conditional unitary routing operation onto positive or negative paths based upon polarization state:
	\begin{equation}
		|L, R\rangle |+, -\rangle \to e^{-i \pm \frac{\pi}{2}} |L, R\rangle |+, -\rangle \tag{7}
	\end{equation}
	This routing operation is directly analogous to the transformation $|w,l\rangle_B |0\rangle_c \mapsto |2w, l+1\rangle_B |0\rangle_c ; |w,l\rangle_B |1\rangle_c \mapsto |2w+1, l+1\rangle_B |1\rangle_c$, as already discussed.
	
	\subsection{Implementation of the QRAM Scheme}
	
	\subsubsection{Routing of the quantum walkers}
	
\begin{figure}[htbp]
		\centering
		\includegraphics[width=0.7\linewidth]{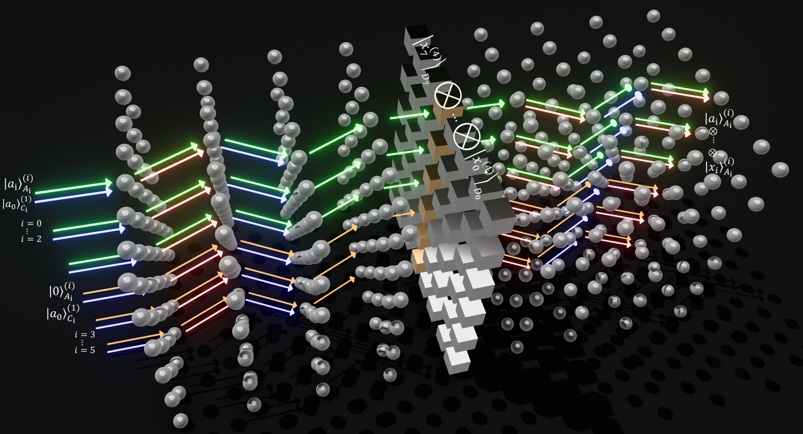}
		\caption{Ensemble of cold alkali atoms (spheres), where each layer serving as a level on a bi-directional binary tree. The address probe laser, signal laser and the data probe laser has been represented by yellow arrows, blue/red arrows for either left/right polarization states and orange arrows. The memory locations, represented by the square arrays.}
		\label{fig:fig3}
	\end{figure}

	To discuss the comprehensive routing scheme, we consider an ensemble of cold alkali atoms, as a collection of cylindrically symmetric hollow-core waveguides, consisting of $N$ atoms each, excited to the Rydberg state, and memory locations, consisting of three-level atomic systems, coupled with a laser field in a Raman transition. Each layer of atoms can be mapped to each level of a binary tree, with the hollow-core waveguides acting as nodes of the binary tree (See Fig.~\ref{fig:fig3}). All the hollow core waveguides are connected by phase-dependent waveguides to amplify the signal of particular phase towards the correct daughter node.
	
\begin{figure}[htbp]
		\centering
		\includegraphics[width=0.7\linewidth]{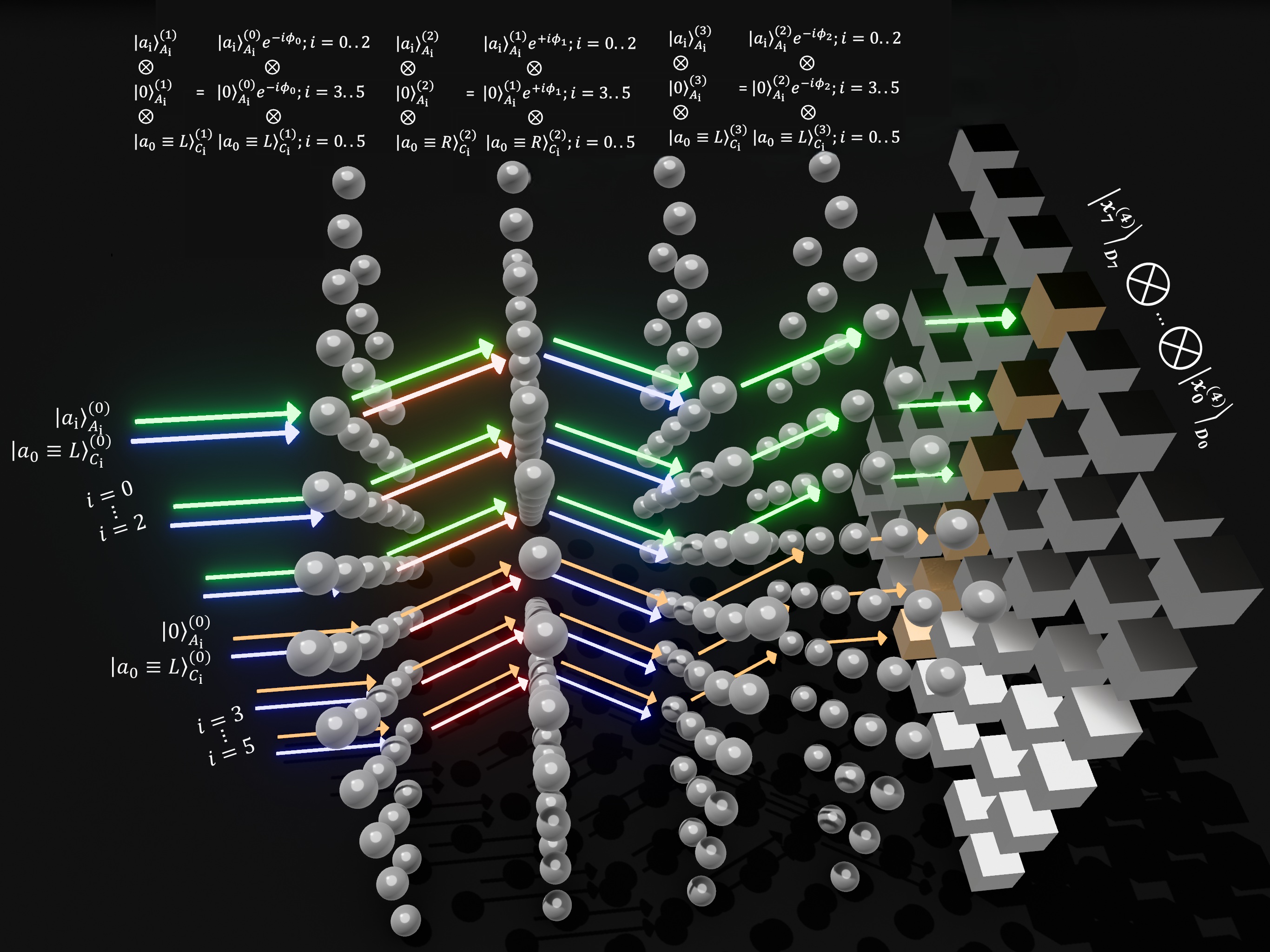}
		\caption{Routing scheme of the quantum walkers, represented by the address and data probe pulses, visualized. Here, shown the routing of probe pulses by ensuing phase generation of $\pm i\phi_n$, based on the left (blue) and right (red) polarization state of signal pulses.}
		\label{fig:fig4}
	\end{figure}

\begin{figure}[htbp]
		\centering
		\includegraphics[width=0.7\linewidth]{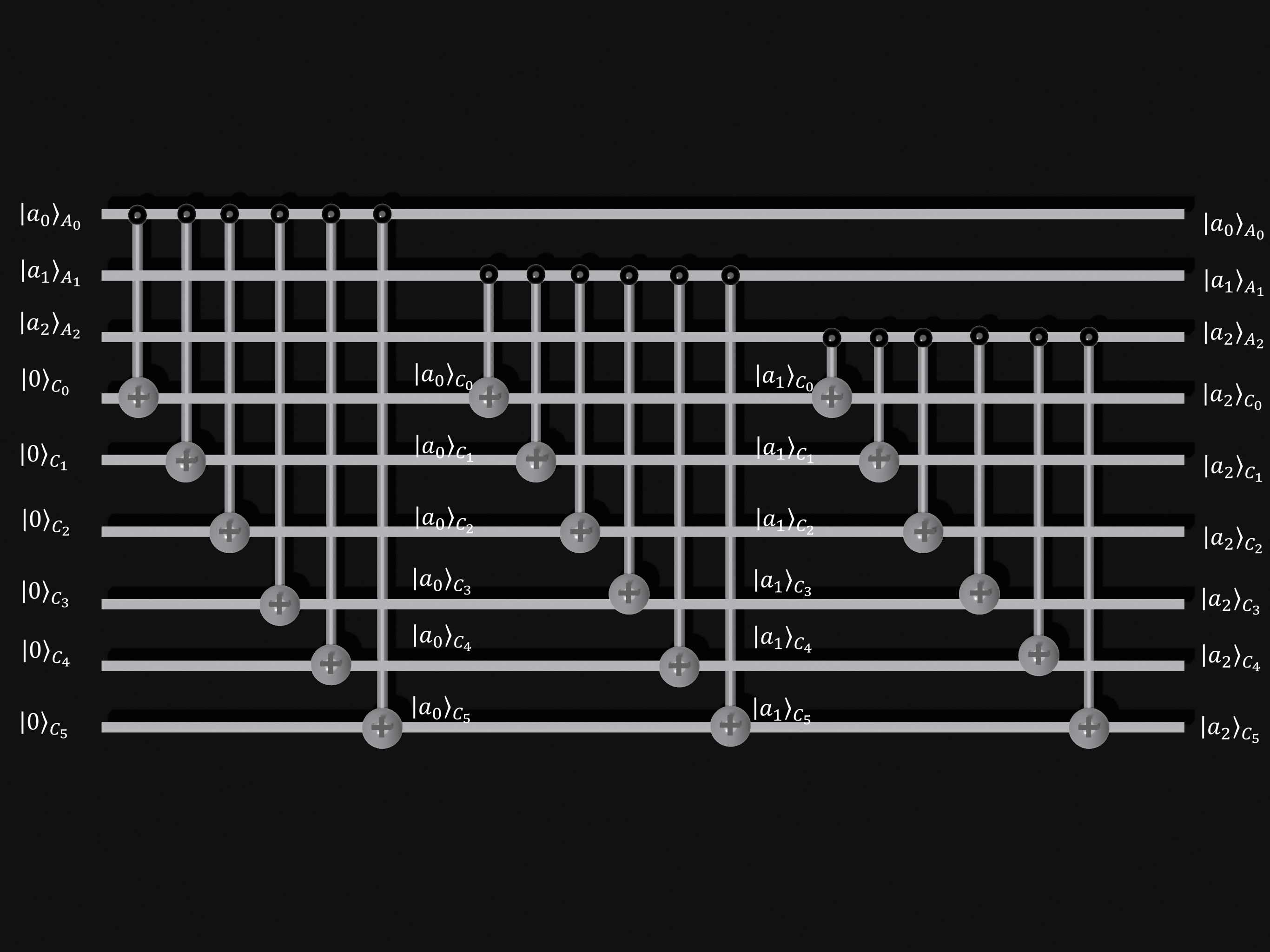}
		\caption{CNOT operations to copy the address information at each level of the binary tree onto the encoded polarization states of the signal laser pulses.}
		\label{fig:fig5}
	\end{figure}

	The routing scheme proceeds as follows (See Fig.~\ref{fig:fig4}). Initially, the $n$-qubit address state information $|\hat{\Psi}_{\text{address}}\rangle = |a_{n-1}\rangle_{A_{n-1}} \otimes \dots \otimes |a_0\rangle_{A_0}$, in the form of polarization states of address probe laser pulses, and the empty $m$-qubit data probe laser pulses $|\hat{\Psi}_{\text{data}}\rangle = |0\rangle_{A_n} \otimes \dots \otimes |0\rangle_{A_{n+m}}$ are initialized. The collective position of all $n + m$ walkers routed among $j$ addresses is $|\hat{\Psi}_{\text{position}}\rangle = |w, \ell\rangle_B = \sum_j |w, \ell\rangle_{B_j}$ where level is $(0 \leqslant w \leqslant n - 1)$ and node is $(0 \leqslant \ell \leqslant 2^w - 1)$. Initially, the address information $|a_{n-1}\rangle_{A_{n-1}} \otimes \dots \otimes |a_0\rangle_{A_0}$ is loaded on all the $n$ address walkers, by encoding them in the polarization state of the $n$ address walker pulses. Beginning at origin node $|a_0\rangle_{A_0}$, address information of subsequent levels $|a_1\rangle_{A_1} \dots |a_{n-1}\rangle_{A_{n-1}}$ is mapped from each of the $n$ address walker probe pulses onto the $n + m$ signal pulses using polarization-dependent optical CNOT gates (See Fig.~\ref{fig:fig5}).
	
	\begin{equation}
		|\Psi_{\text{signal}}\rangle = (|0\rangle_{c_{n+m}} \otimes \dots \otimes |0\rangle_{c_0}) \xrightarrow{\text{CNOT}_{a_l A_l}} |a_l\rangle_{c_{n+m}} \otimes \dots \otimes |a_l\rangle_{c_0} \quad \text{for } l^{\text{th}} \text{ level of the binary tree} \tag{8}
	\end{equation}
	
	Node address mapping employs a polarization-dependent optical CNOT gate on an integrated optics platform using an entangled Bell state $|\Psi^-\rangle = (|H, V\rangle - |V, H\rangle)/\sqrt{2}$. Birefringent beam splitters retain vertical and transfer horizontal light. Operating probabilistically (1/4 success), it is heralded by coincidence measurements, followed by feedforward Pauli unitaries. Minimizing optical losses coherently encodes quantum walker address states onto signal pulses for deterministic routing of Rydberg ensembles \cite{zeuner2018}.
	
	Storing addresses via $|0\rangle_c \to |a_{n-1}\rangle_c$, qubits $|0\rangle, |1\rangle$ encode the walker probes' polarizations $|L\rangle, |R\rangle$ onto signal probes, modifying the signal field's number excitation operator $\hat{I}_{l'}(z', t)$. The $n + m$ probes couple the ground state $|g\rangle$ and excited states $|e_{1,2}\rangle$ using principal quantum numbers $n = 10$ and $n = 15$ to maintain robust electromagnetically induced transparency (EIT) without spectral degradation. Dark state polaritons undergo cross-phase modulation via strong dipole-dipole interactions and acquire conditional phase shifts. Mach-Zehnder interferometers \cite{xie2020, zheng2023} spatially resolve these shifts via the phase difference $\Delta\varphi$. Modulating the waveguide coupling deterministically routes the retrieved walkers to the left ($|w + 1, 2\ell\rangle_{B_i}$) or right ($|w + 1, 2\ell + 1\rangle_{B_i}$) daughter nodes, obeying Eq. (7).
	
	\begin{align}
		|L, R\rangle_{C_i} |a_i\rangle_{A_i} |w, \ell\rangle_{B_i} &\to e^{+i\phi_n} |L, R\rangle_C |a_i\rangle_{A_i} |w + 1, 2\ell\rangle_{B_i} \nonumber \\
		|L, R\rangle_{C_i} |a_i\rangle_{A_i} |w, \ell\rangle_{B_i} &\to e^{-i\phi_n} |L, R\rangle_C |a_i\rangle_{A_i} |w + 1, 2\ell + 1\rangle_{B_i} \tag{9}
	\end{align}
	
	The $n + m$ signal pulses are retrieved in the opposite direction, enabling next address qubit encoding. Likewise, the routing process continues until the incoming walker probe pulses $|\hat{\Psi}_{\text{input}}\rangle = |\hat{\Psi}_{\text{address}}\rangle \otimes |\hat{\Psi}_{\text{data}}\rangle \otimes |\hat{\Psi}_{\text{position}}\rangle \otimes |\hat{\Psi}_{\text{signal}}\rangle$ reach the designated memory location. After the routing process concludes, all $n + m$ walker probe pulses acquire cumulative phase as they approach the $k\text{th}$ memory location at the $n\text{th}$ level:
	\begin{align}
		&(|a_{n-1}\rangle_{A_{n-1}} \otimes \cdots \otimes |a_0\rangle_{A_0}) \otimes (|0\rangle_{A_n} \otimes \cdots \otimes |0\rangle_{A_{n+m}} \otimes |a_l\rangle_{c_{n+m}}) \otimes_i |0,0\rangle_{B_i} \otimes (|a_l\rangle_{c_{n+m}} \otimes \cdots \otimes |a_l\rangle_{c_0}) \nonumber \\
		&\xrightarrow{\textit{Routing}} \nonumber \\
		&\{\exp[-i\phi_0] \dots \exp[-i\phi_n]\} \times (|a_{n-1}\rangle_{A_{n-1}} \otimes \cdots \otimes |a_0\rangle_{A_0}) \otimes (|0\rangle_{A_n} \otimes \cdots \otimes |0\rangle_{A_{n+m}}) \bigotimes_{i=0}^{i=n+m} |n, k\rangle_{B_i} \otimes \nonumber \\
		&\qquad\qquad (|a_n\rangle_{c_{n+m}} \otimes \cdots \otimes |a_n\rangle_{c_0}) \tag{10}
	\end{align}
	Here, $\phi_n$ is the phase accumulated at the $n^{\text{th}}$ level of the binary tree in each layer by each address walker pulse.
	\FloatBarrier
	\subsubsection{Querying for the data}
\begin{figure}[htbp]
		\centering
		\includegraphics[width=0.7\linewidth]{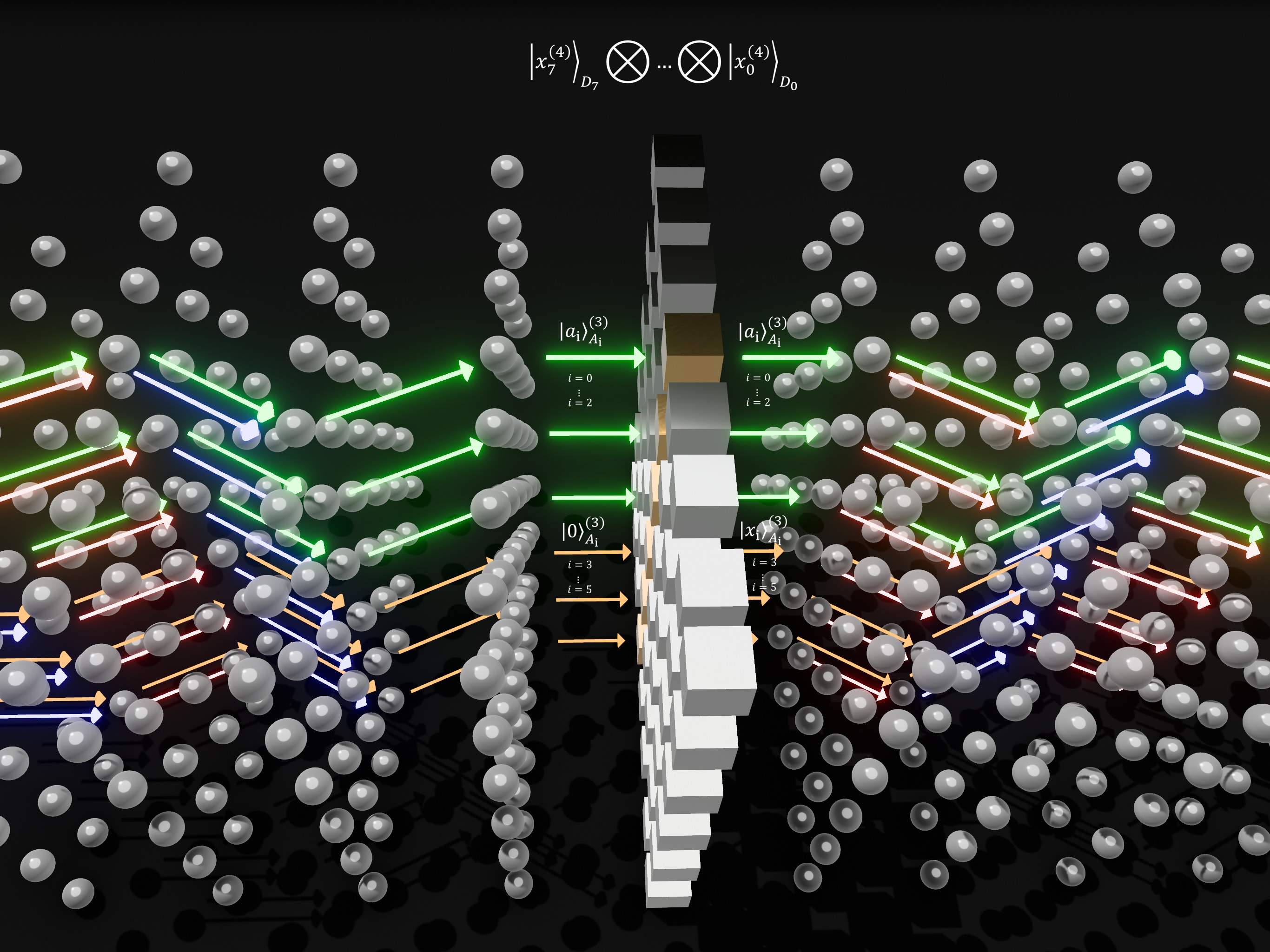}
		\caption{Data querying through memory locations.}
		\label{fig:fig6}
	\end{figure}

	Each memory location is a three-level atomic system (degenerate ground states $|g\rangle, |e\rangle$; excited state $|r\rangle$), storing data as a superposition $|\psi\rangle = \alpha|g\rangle + \beta|e\rangle$. By mapping the state of a transmitting atom into a photonic wave packet, sculpted driving pulses emit a time-symmetric photon with unit efficiency and without reflection loss. The receiving atom mimics the time-reversed emission by absorbing it with unit probability to restore the superposition. When $n + m$ walkers arrive, $m$ data walkers in the collective Fock states $|0\rangle_{A_n} \otimes \dots \otimes |0\rangle_{A_{n+m}}$ generate cavity modes $\hat{a}_i$ (See Fig.~\ref{fig:fig6}).
	
	During querying, an external laser $\omega_L$ couples $|e\rangle$ to $|r\rangle$ (excitation frequency $\omega_0$) via a Raman transition parameterized by the Rabi frequency $\Omega_i(t)$ and phase $\phi_i(t)$. Simultaneously, atom-cavity coupling $g$ governs the $|r\rangle \to |g\rangle$ transition, emitting or absorbing photons. These dynamics, along with the cavity mode (destruction operator $\hat{a}_i$, resonant frequency $\omega_c$), define the time-dependent Hamiltonian of the $i\text{th}$ node \cite{cirac1997}:
	\begin{equation}
		\widehat{H}_i = \omega_c \hat{a}_i^\dagger \hat{a}_i + \omega_0 |r\rangle_{ii}\langle r| + g(|r\rangle_{ii}\langle g|\hat{a}_i + \text{h.c.}) + \frac{1}{2}\Omega_i(t) \left[ e^{-i[\omega_L t + \phi_i(t)]} |r\rangle_{ii}\langle e| + \text{h.c.} \right] (i = 1,2) \tag{11}
	\end{equation}
	
\begin{figure}[htbp]
		\centering
		\includegraphics[width=0.7\linewidth]{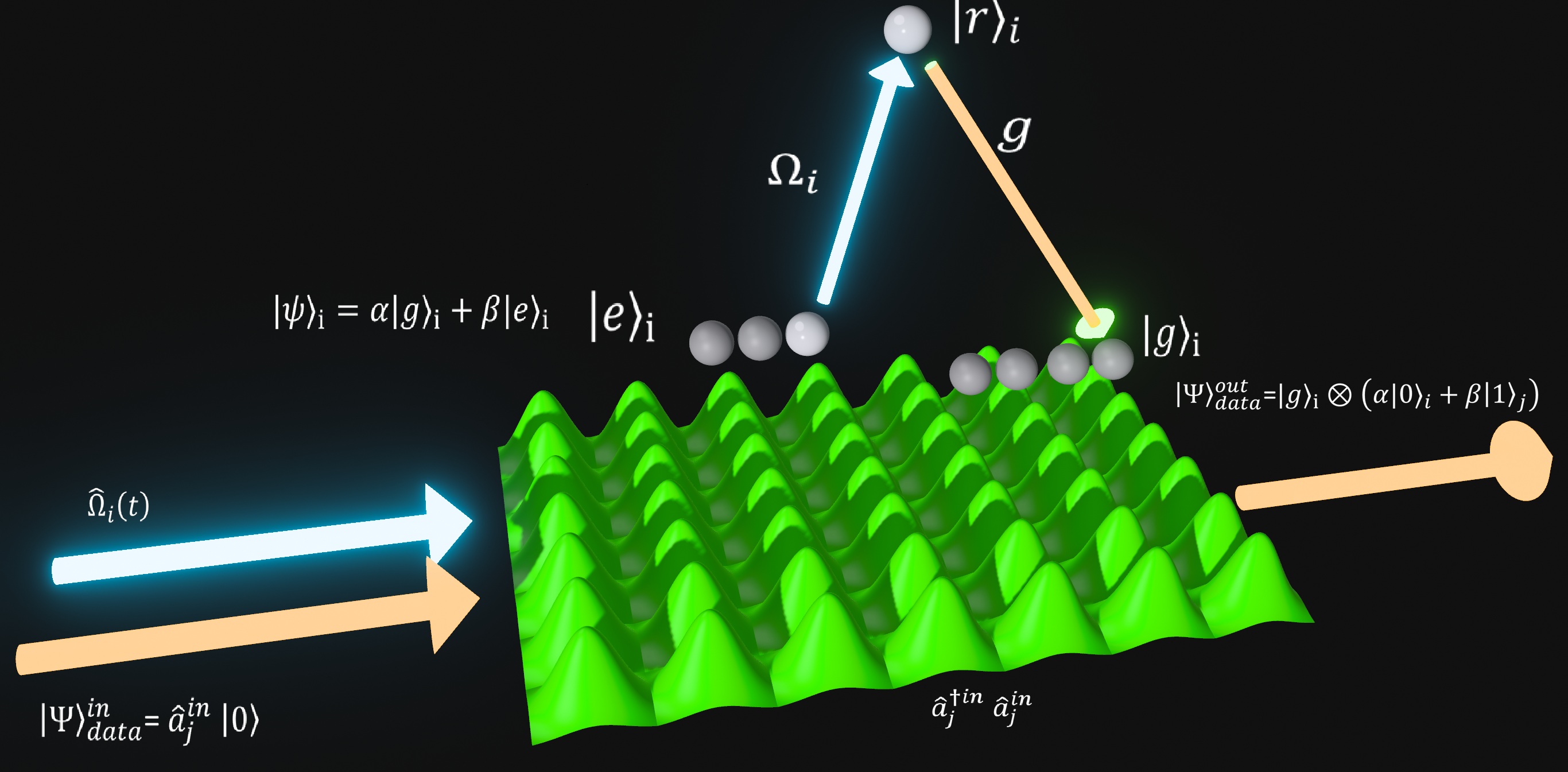}
		\caption{Three-level Raman transition and mapping of stored data in the form of superposition of atomic states onto the data probe laser.}
		\label{fig:fig7}
	\end{figure}

	Experimentally, this querying architecture is realized by dropping atoms from a magneto-optical trap and cooling a single cesium atom into a far-off-resonant trap (FORT) within the cavity mode, placing the system deeply in the strong-coupling regime of cavity quantum electrodynamics. Within this setup, the memory location stores data as a coherent ground-state superposition defined as $|\psi\rangle = \alpha|g\rangle + \beta|e\rangle$, which corresponds physically to the $F = 3$ and $F = 4$ hyperfine ground states of the Cesium atom \cite{boozer2007}. To query the data, the experimental sequence dictates applying a precisely timed classical laser pulse $\Omega_i(t)$ transverse to the cavity axis, which initiates a dark-state adiabatic passage that transfers the atomic population and stimulates the coherent emission of a corresponding single-photon field inside the cavity. After emission, the atom is left securely in the ground state $|g\rangle_i$, while the emitted photon, which escapes through the cavity output mirror and is detected by a single-photon counting avalanche photodiodes, coherently encodes the quantum information of the original qubit (See Fig.~\ref{fig:fig7}). This reversible mapping translates the atomic superposition into the presence or absence of a photon within a well-defined wave packet model:
	\begin{equation}
		|\psi\rangle_i \otimes |\hat{\Psi}_{\text{data}}\rangle_j = (\alpha|g\rangle_i + \beta|e\rangle_i) \otimes |0\rangle_j \to |g\rangle_i \otimes (\alpha|0\rangle_j + \beta|1\rangle_j) \tag{12}
	\end{equation}
	where: $|0\rangle_j$ is the $j\text{th}$ vacuum state and $|1\rangle_j = \int dt \xi(t) \hat{a}_{\text{out}}^\dagger(t) |0\rangle_j$ is the single photon state. In this transformation, $|0\rangle_j$ represents the $j\text{th}$ vacuum state, and $|1\rangle_j = \int dt \xi(t) \hat{a}_{\text{out}}^\dagger(t) |0\rangle_j$ is the emitted single-photon state. This process faithfully maps the atomic qubit's amplitudes onto the photonic wave packet, where the wave packet's temporal profile $\xi(t)$ is strictly determined by the temporal envelope of the Raman control pulse. By symmetrically shaping this control pulse, the generated wave packet can subsequently be absorbed by a receiving cavity with unit efficiency. The coherence of this querying step is explicitly verified experimentally by combining the emitted output field with a reference phase-coherent laser pulse and observing the resulting visibility fringes \cite{boozer2007}.
	
	Once the state is mapped onto this generated light field, the photon exits the cavity through an output mirror and is directed to a pair of single-photon counting avalanche photodiodes. These detectors are used to explicitly read the state stored in the light pulse. To verify that the coherent superposition of the original atomic state was faithfully preserved and mapped onto the photon, the emitted field is combined with a reference light pulse to create interference. By observing the photon counts at the avalanche photodiodes as a function of the relative phase between the initial and reference fields. This direct detection via the photodiodes successfully confirmed the coherent transfer of the quantum state from the single trapped atom directly to the emitted light pulse \cite{boozer2007}. Thus, the data states $|x^{(a)}\rangle_D = \left|x_{m-1}^{(a)}\right\rangle_{D_{m-1}} \otimes \dots \otimes \left|x_0^{(a)}\right\rangle_{D_0} = (\alpha|0\rangle_\gamma + \beta|1_\xi\rangle_\gamma)_{m-1} \otimes \dots \otimes (\alpha|0\rangle_\gamma + \beta|1_\xi\rangle_\gamma)_0$ in the form of the long-lived atomic states $|\psi\rangle_1 \otimes \dots \otimes |\psi\rangle_m$ is mapped onto the $m$ probe walker states as:
	\begin{align}
		&\bigotimes_{i=0}^{m-1} (\alpha|g\rangle_i + \beta|e\rangle_i) \otimes (|0\rangle_{A_n} \otimes \dots \otimes |0\rangle_{A_{n+m}}) \to \bigotimes_{i=n}^{n+m} |g\rangle_i \otimes (\alpha|0\rangle_\gamma + \beta|1_\xi\rangle_\gamma)_{m-1} \otimes \dots \otimes (\alpha|0\rangle_\gamma + \beta|1_\xi\rangle_\gamma)_0 \nonumber \\
		&\text{or} \nonumber \\
		&\bigotimes_{i=0}^{m-1} (\alpha|g\rangle_i + \beta|e\rangle_i) \otimes (|0\rangle_{A_n} \otimes \dots \otimes |0\rangle_{A_{n+m}}) \to \bigotimes_{i=n}^{n+m} |g\rangle_i \otimes \left(|x_0^{(a)}\rangle_{A_n} \otimes \dots \otimes |x_{m-1}^{(a)}\rangle_{A_{n+m}}\right) \tag{13}
	\end{align}
	for each of the $m$ data qubits $\left|x_{m-1}^{(a)}\right\rangle_{D_{m-1}} \otimes \dots \otimes \left|x_0^{(a)}\right\rangle_{D_0}$.
	
	Such that the queried collective quantum walker state becomes
	\begin{align}
		&(|a_{n-1}\rangle_{A_{n-1}} \otimes \cdots \otimes |a_0\rangle_{A_0}) \otimes (|0\rangle_{A_n} \otimes \cdots \otimes |0\rangle_{A_{n+m}} \otimes |a_l\rangle_{c_{n+m}}) \otimes_i |0,0\rangle_{B_i} \otimes (|a_l\rangle_{c_{n+m}} \otimes \cdots \otimes |a_l\rangle_{c_0}) \nonumber \\
		&\xrightarrow{\textit{Routing}} \nonumber \\
		&\{\exp[-i\phi_0] \dots \exp[-i\phi_n]\} \times (|a_{n-1}\rangle_{A_{n-1}} \otimes \cdots \otimes |a_0\rangle_{A_0}) \otimes \left(|x_0^{(a)}\rangle_{A_n} \otimes \cdots \otimes |x_{m-1}^{(a)}\rangle_{A_{n+m}}\right) \bigotimes_{i=0}^{i=n+m} |n, k\rangle_{B_i} \nonumber \\
		&\qquad\qquad \otimes (|a_n\rangle_{c_{n+m}} \otimes \cdots \otimes |a_n\rangle_{c_0}) \tag{14}
	\end{align}
	
	\subsubsection{Retrieving the data}
	
\begin{figure}[htbp]
		\centering
		\includegraphics[width=0.7\linewidth]{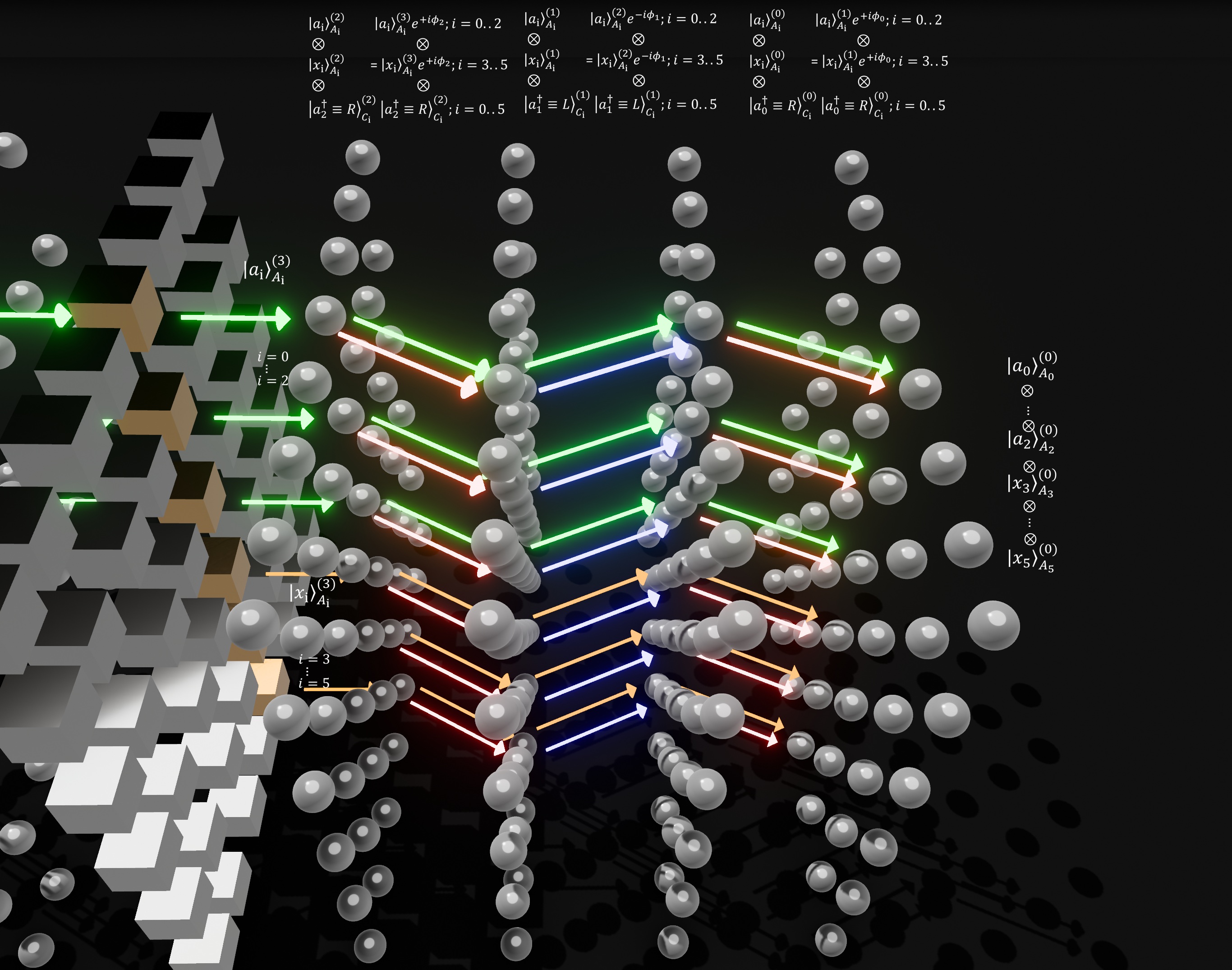}
		\caption{Data retrieval through reverse binary tree.}
		\label{fig:fig8}
	\end{figure}

	After the data in the memory location are mapped into the $m$ data walkers' fock states, the $n+m$ walker pulses are directed towards the inverted binary tree in each of the $n + m$ layers. The $n + m$ walker pulses are therefore first directed towards the $n - 1\text{th}$ level of the inverted binary tree, consisting of the exactly same nodes, just this time, inverse operations to those have been discussed under the routing scheme are performed (See Fig.~\ref{fig:fig8}).
	Initially, the address state information of each qubit $|a_{n-1}\rangle_{A_{n-1}} \otimes \dots \otimes |a_0\rangle_{A_0}$ is encoded as the polarization state in the address state walkers, just this time the opposite state to the address associated with the $(n - l)\text{th}$ level of the tree $|a_{n-l}\rangle_{A_{n-l}}$ is encoded into the polarization state of the $n + m$ signal pulses, i.e. $|a_n\rangle_{A_n}$ address qubit is encoded into the $n + m$ signal pulses at $n\text{th}$ level of the binary tree
	\begin{align}
		|\Psi_{\text{signal}}\rangle = (|0\rangle_{c_{n+m}} \otimes \dots \otimes |0\rangle_{c_0}) . CNOT . |a_l\rangle_{A_l} \to |a_l\rangle_{c_{n+m}} \otimes \dots \otimes |a_l\rangle_{c_0} \nonumber \\
		\text{for } (n - l)\text{th level of the binary tree} \tag{15}
	\end{align}
	The $n + m$ signal and walker probes are then made incident on the particular node and undergo photon-photon cross-phase modulation, directing the retrieved walker probes traveling in positive or negative waveguide directions towards the left (with odd index $|w + 1, 2\ell + 1\rangle_{B_i}$) or right daughter node(with even index $|w + 1, 2\ell\rangle_{B_i}$), obeying the opposite of the routing procedure in Eq. (9).
	\begin{equation}
		\left.
		\begin{aligned}
			|L, R\rangle_{C_i} |a_i\rangle_{A_i} |w, \ell\rangle_{B_i} &\to e^{-i\phi_n} |L, R\rangle_C |a_i\rangle_{A_i} |w + 1, 2\ell\rangle_{B_i} \\
			|L, R\rangle_{C_i} |a_i\rangle_{A_i} |w, \ell\rangle_{B_i} &\to e^{+i\phi_n} |L, R\rangle_C |a_i\rangle_{A_i} |w + 1, 2\ell + 1\rangle_{B_i}
		\end{aligned}
		\right\} \tag{16}
	\end{equation}
	The signal and walker probe pulses are made incident on the particular node of the $n - 1\text{th}$ layer, and after the cross-phase modulation due to the signal pulse-walker pulse interaction at the node, the walker probe pulse is retrieved, and the unitary of the operation is conserved as the phase $-i\phi_l$ the walkers gained at $l\text{th}$ level of the binary tree during the routing is cancelled by the exactly opposite phase $+i\phi_l$ it gained during the retrieval process at the $l\text{th}$ level, such that after $n - 1$ steps, we obtain:
	\begin{align}
		&\{\exp[-i\phi_0] \dots \exp[-i\phi_n]\} \times (|a_{n-1}\rangle_{A_{n-1}} \otimes \cdots \otimes |a_0\rangle_{A_0}) \otimes \left(|x_0^{(a)}\rangle_{A_n} \otimes \cdots \otimes |x_{m-1}^{(a)}\rangle_{A_{n+m}}\right) \bigotimes_{i=0}^{i=n+m} |n, k\rangle_{B_i} \nonumber \\
		&\qquad \otimes (|a_n\rangle_{c_{n+m}} \otimes \cdots \otimes |a_n\rangle_{c_0}) \xrightarrow{\textit{Routing}} \nonumber \\
		&\{\exp[+i\phi_0] \dots \exp[+i\phi_n]\} \times \{\exp[-i\phi_0] \dots \exp[-i\phi_n]\} \times (|a_{n-1}\rangle_{A_{n-1}} \otimes \cdots \otimes |a_0\rangle_{A_0}) \nonumber \\
		&\qquad \otimes \left(|x_0^{(a)}\rangle_{A_n} \otimes \cdots \otimes |x_{m-1}^{(a)}\rangle_{A_{n+m}}\right) \bigotimes_{i=0}^{i=n+m} |0,0\rangle_{B_i} \otimes (|a_0\rangle_{c_{n+m}} \otimes \cdots \otimes |a_0\rangle_{c_0}) \nonumber \\
		&\Rightarrow (|a_{n-1}\rangle_{A_{n-1}} \otimes \cdots \otimes |a_0\rangle_{A_0}) \otimes \left(|x_0^{(a)}\rangle_{A_n} \otimes \cdots \otimes |x_{m-1}^{(a)}\rangle_{A_{n+m}}\right) \bigotimes_{i=0}^{i=n+m} |n, k\rangle_{B_i} \otimes (|a_n\rangle_{c_{n+m}} \otimes \cdots \otimes |a_n\rangle_{c_0}) \nonumber \\
		&\xrightarrow{\textit{Routing}} (|a_{n-1}\rangle_{A_{n-1}} \otimes \cdots \otimes |a_0\rangle_{A_0}) \otimes \left(|x_0^{(a)}\rangle_{A_n} \otimes \cdots \otimes |x_{m-1}^{(a)}\rangle_{A_{n+m}}\right) \bigotimes_{i=0}^{i=n+m} |0,0\rangle_{B_i} \otimes (|a_0\rangle_{c_{n+m}} \otimes \cdots \otimes |a_0\rangle_{c_0}) \tag{17}
	\end{align}
	As can be seen, the total routing to retrieval of the quantum address state walker and the data state walker therefore satisfy the condition of the QRAM, obeying the relation :
	\begin{align*}
		&|a_{n-1}\rangle_{A_{n-1}} \otimes \dots \otimes |a_0\rangle_{A_0} \otimes |0\rangle_{A_n} \otimes \dots \otimes |0\rangle_{A_{n+m}} \bigotimes_{i=0}^{i=n+m} |0,0\rangle_{B_i} \\
		&\to |a_{n-1}\rangle_{A_{n-1}} \otimes \dots \otimes |a_0\rangle_{A_0} \otimes \left|x_{m-1}^{(a)}\right\rangle_{A_n} \otimes \dots \otimes \left|x_0^{(a)}\right\rangle_{A_{n+m}} \bigotimes_{i=0}^{i=n+m} |0,0\rangle_{B_i}
	\end{align*} 
	
	\section{Results and discussions}
	
	The experimental realization of selective phase generation must satisfy three conditions outlined in \cite{shahmoon2011, petrosyan2012, friedler2005_2, zeuner2018}: 
	i) The duration $\tau$ of the address and signal probe pulses must exceed the inverse EIT bandwidth $\delta\omega_l = |\Omega_l|^2/(\gamma_{ge_l}\sqrt{\kappa_l L})$: $T > (\gamma_{ge_l}\sqrt{\kappa_l L})/|\Omega_l|^2$, where $\gamma_{ge_l}$ is the ground-state relaxation rate and $\kappa_l L$ is the absorption coefficient, requiring group velocity $v_g$ to satisfy $\tau \gg \Delta\omega_{\text{EIT}}^{-1}$ \cite{friedler2005_2, zeuner2018}. 
	ii) The total dipole-dipole induced phase shift $2\phi_{ll'} \tilde{n}_{l'}$ should remain below the EIT bandwidth $\Delta\omega_{\text{EIT}}$, meaning $\delta\omega_l = |\Omega_l|^2/(\gamma_{ge_l}\sqrt{\kappa_l L})$: $2\phi_{ll'} \tilde{n}_{l'} < |\Omega_l|^2/(\gamma_{ge_l}\sqrt{\kappa_l L})$ for average photon number $\tilde{n}_s$ and temporal length $\tau$. 
	iii) The propagation time $t_{\text{prop}} = L/v_g$ inside a waveguide of length $L$ is limited by the relaxation rate $\gamma_{gd_l}$ \cite{shahmoon2011}. 
	
	\begin{figure}[htbp]
		\centering
		\includegraphics[width=0.7\linewidth]{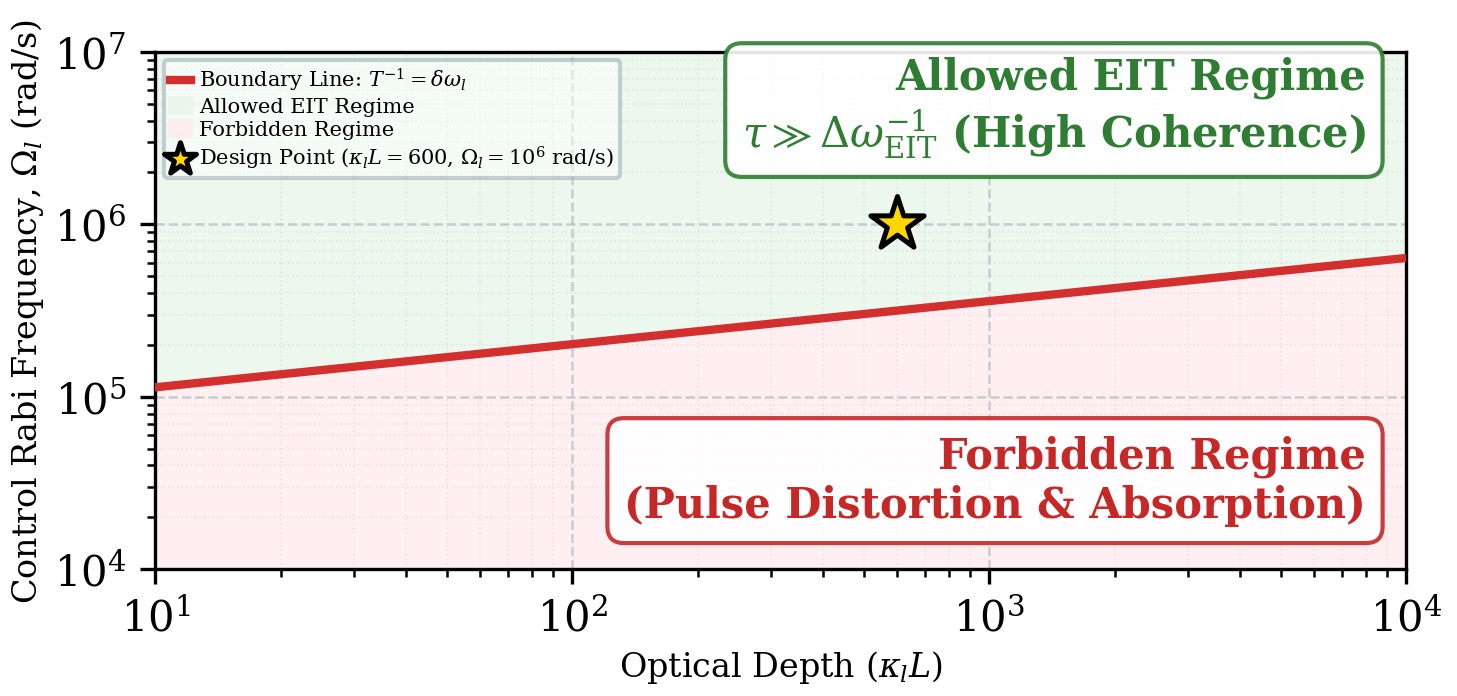}
		\caption{Operational parameter space for dark-state polariton quantum walkers. Control Rabi frequency $\Omega_l$ is plotted against optical depth $\kappa_l L$ on logarithmic scales to validate condition (i).}
		\label{fig:fig9}
	\end{figure}
	
\begin{figure}[htbp]
		\centering
		\includegraphics[width=0.7\linewidth]{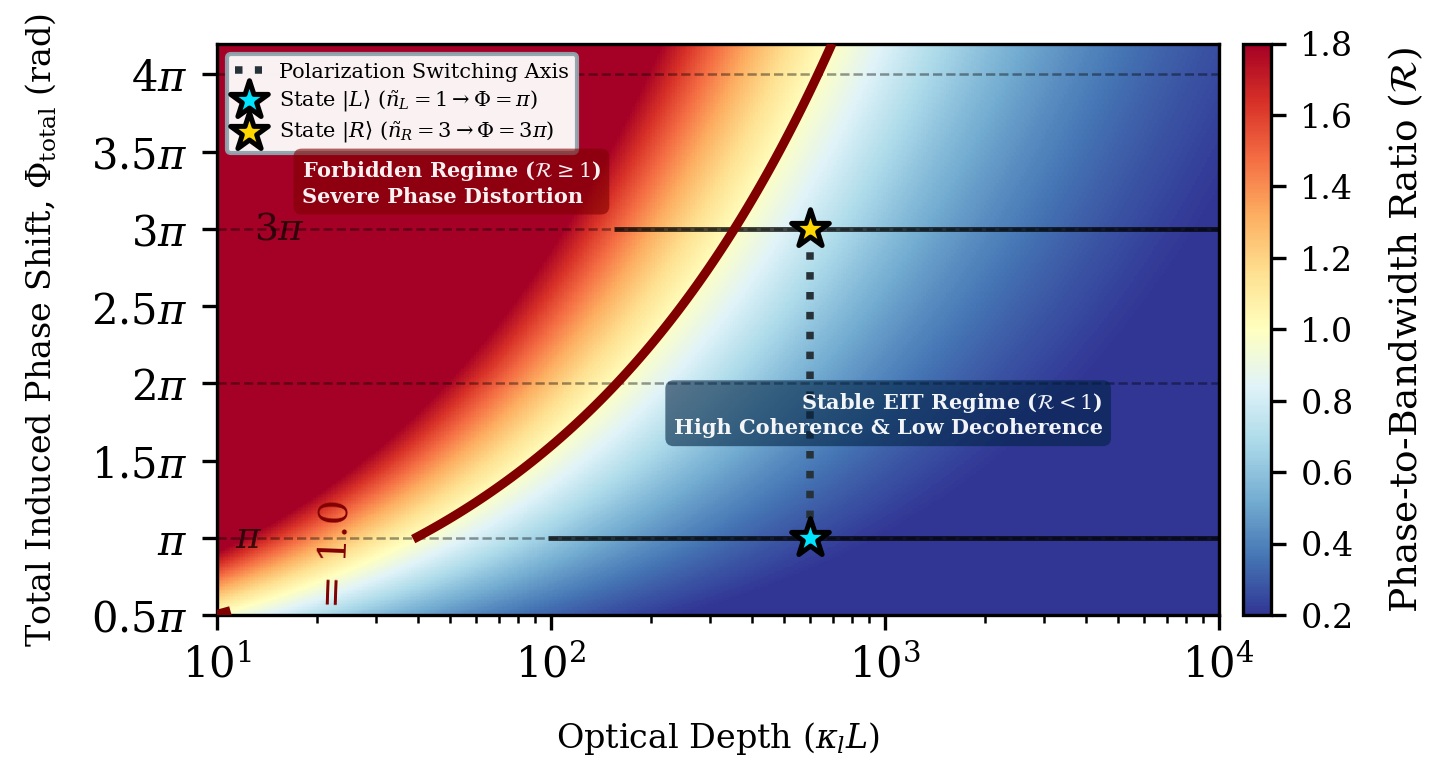}
		\caption{Cross-phase modulation stability heatmap and differential phase routing in the Rydberg-EIT qRAM. The color map shows the phase-to-bandwidth ratio $\mathcal{R} = 2\phi_{ll}n_l / \delta\omega_l$ across optical depth $\kappa_l L$ and conditional phase shift $\Phi_{\text{total}}$.}
		\label{fig:fig10}
	\end{figure}

	We assume atom number $N = 10^4$, optical depth $\kappa_l L \sim 600$, control Rabi frequencies $\Omega = 10^6\text{ rad/s}$, group velocity $v_g = 100\text{ m/s}$, pulse bandwidth $T^{-1} \sim 10^4$, waveguide length $L = 1\text{ cm}$, transverse Gaussian width $w_f \sim 2\ \mu\text{m}$, and atomic cavity width $w_a \sim 2\ \mu\text{m}$. These parameters perfectly satisfy conditions i) and ii), because pulse bandwidth is less than EIT bandwidth, $T^{-1} \sim 10^4 < \delta\omega_l \sim 10^5$, and for conditional phase shifts $\phi_{ll'}\tilde{n}_l \approx \frac{\pi}{2}$ and $\frac{3\pi}{2}$, the total maximum phase shift remains $2\phi_{ll'}\tilde{n}_l = 3\pi < \frac{1}{2}\sqrt{\kappa_l L} \sim 12.25$. To implement this routing procedure, we select Rydberg states with dipole moments $\wp_{d_l} = 150 e a_0$ having quantum numbers $(n = 10, l = 0)$ and $\wp_{d_p} = 337.5 e a_0$ having quantum numbers $(n = 10, l = 0)$ and $(n = 15, l = 0)$ for address and signal walker probes, respectively (See Fig.~\ref{fig:fig9} and Fig.~\ref{fig:fig10}).
	
	Unlike conventional fanout's exponential gate overhead and bucket brigade's $\mathcal{O}(n^2)$ error scaling, this scheme eliminates the maintenance of coherence across $n \times 2^n$ active qutrit switches by utilizing an internal chirality-driven unitary quantum random walk, thereby reducing access complexity to $\mathcal{O}(n)$. Instead of toggling active physical gates, $n$ address polarization states transfer directly to auxiliary signal pulses. Walkers accumulate conditional phase shifts via electromagnetically induced transparency (EIT) and strong Rydberg dipole-dipole interactions, executing exactly $n$ sequential phase-dependent directional shifts rather than geometric phase gates. Traversing from the root of the tree to any specific memory location therefore requires exactly $n$ sequential unitary operations. By replacing localized node switching with an inherent state-driven traversal across the $n$ levels, the computational access complexity is mathematically strictly linear, equating to $\mathcal{O}(n)$. Furthermore, the cumulative phase equations confirm that this sequential scaling ensures linear phase accumulation relative to the tree depth, guaranteeing predictable routing with vastly diminished decoherence.

	Querying maps metastable atomic qubits to photonic wave packets via cavity-assisted Raman processes, faithfully preserving superposition amplitudes for stationary-to-flying qubit transfer. Retrieval reversibility via inverse tree operations disentangles probes to establish the necessary protocol unitarity. Guided photonic qubits offer stronger isolation and minimal dephasing compared to cryogenic platforms. Despite requiring precise laser detuning, cavity-waveguide loss mitigation, and parallel synchronization, integrating multilevel EIT with Rydberg interactions merges photonic speed and atomic coherence for a scalable, programmable qRAM.
	
	\section{Conclusion}
	We proposed a novel quantum random access memory (qRAM) architecture harnesses quantum random walks within a Rydberg-blockaded atomic ensemble mediated by multilevel electromagnetically induced transparency (EIT). By encoding address information in photonic polarization states and routing through a binary tree, conditional cross-phase modulation achieves deterministic routing with an operational complexity of $\mathcal{O}(n)$. Querying and retrieval processes confirm protocol reversibility and unitarity, enabling coherent stationary-to-flying qubit transfer. This approach eliminates exponential fanout gate overhead, leverages robust Rydberg dipole-dipole interactions, and provides high-fidelity non-cryogenic mapping. Despite experimental challenges regarding EIT resonance stabilization, cavity-waveguide photon loss, and multi-channel synchronization, our framework establishes a scalable pathway for programmable quantum memory, serving as an essential cornerstone for future quantum computing, communication, and artificial intelligence architectures.
	
	\section*{Acknowledgement}
	Authors would like to thank their host institutions and their respective funding agencies.
	
	\section*{Data Availability}
	Not Applicable.
	
	\section*{Conflict of Interest}
	The authors have no relevant financial or non-financial interests to disclose.
	
	\section*{Authors’ Contribution}
	All authors contributed equally.
	
	\bibliographystyle{apsrev4-2} 
	\bibliography{Quantum_RAM_Implementation_Using_Multiple_Interacting_Rydberg_Blockaded_EIT_Systems} 

@article{benioff1980,
	title={The computer as a physical system: A microscopic quantum mechanical Hamiltonian model of computers as represented by Turing machines},
	author={Benioff, P.},
	journal={Journal of Statistical Physics},
	volume={22},
	pages={563--591},
	year={1980}
}

@article{feynman1982,
	title={Simulating physics with computers},
	author={Feynman, R. P.},
	journal={International Journal of Theoretical Physics},
	volume={21},
	pages={467--488},
	year={1982}
}

@article{deutsch1985,
	title={Quantum theory, the Church--Turing principle and the universal quantum computer},
	author={Deutsch, D.},
	journal={Proceedings of the Royal Society of London. A. Mathematical and Physical Sciences},
	volume={400},
	pages={97--117},
	year={1985}
}

@inproceedings{bernstein1993,
	title={Quantum complexity theory},
	author={Bernstein, E. and Vazirani, U.},
	booktitle={Proceedings of the twenty-fifth annual ACM symposium on Theory of Computing, San Diego, California, USA},
	year={1993}
}

@inproceedings{berthiaume1992,
	title={The quantum challenge to structural complexity theory},
	author={Berthiaume, A. and Brassard, G.},
	booktitle={Proceedings of the Seventh Annual Structure in Complexity Theory Conference},
	pages={132--137},
	year={1992}
}

@inproceedings{shor1994,
	title={Algorithms for quantum computation: discrete logarithms and factoring},
	author={Shor, P. W.},
	booktitle={Proceedings 35th Annual Symposium on Foundations of Computer Science},
	pages={124--134},
	year={1994}
}

@inproceedings{grover1996,
	title={A fast quantum mechanical algorithm for database search},
	author={Grover, L. K.},
	booktitle={Proceedings of the twenty-eighth annual ACM symposium on Theory of Computing, Philadelphia},
	year={1996}
}

@article{harrow2009,
	title={Quantum Algorithm for Linear Systems of Equations},
	author={Harrow, A. W. and Hassidim, A. and Lloyd, S.},
	journal={Physical Review Letters},
	volume={103},
	pages={150502},
	year={2009}
}

@misc{lloyd2013,
	title={Quantum algorithms for supervised and unsupervised machine learning},
	author={Lloyd, S. and Mohseni, M. and Rebentrost, P.},
	howpublished={arXiv: Quantum Physics},
	year={2013}
}

@article{aaronson2015,
	title={Read the fine print},
	author={Aaronson, S.},
	journal={Nature Physics},
	volume={11},
	pages={291--293},
	year={2015}
}

@article{biamonte2017,
	title={Quantum machine learning},
	author={Biamonte, J. and Wittek, P. and Pancotti, N. and Rebentrost, P. and Wiebe, N. and Lloyd, S.},
	journal={Nature},
	volume={549},
	pages={195--202},
	year={2017}
}

@article{giovannetti2008_1,
	title={Architectures for a quantum random access memory},
	author={Giovannetti, V. and Lloyd, S. and Maccone, L.},
	journal={Physical Review A},
	volume={78},
	pages={052310},
	year={2008}
}

@article{giovannetti2008_2,
	title={Quantum Private Queries},
	author={Giovannetti, V. and Lloyd, S. and Maccone, L.},
	journal={Physical Review Letters},
	volume={100},
	pages={230502},
	year={2008}
}

@article{giovannetti2008_3,
	title={Quantum Random Access Memory},
	author={Giovannetti, V. and Lloyd, S. and Maccone, L.},
	journal={Physical Review Letters},
	volume={100},
	pages={160501},
	year={2008}
}

@article{duan2003,
	title={Controlling Spin Exchange Interactions of Ultracold Atoms in Optical Lattices},
	author={Duan, L. M. and Demler, E. and Lukin, M. D.},
	journal={Physical Review Letters},
	volume={91},
	pages={090402},
	year={2003}
}

@article{farhi1998,
	title={Quantum computation and decision trees},
	author={Farhi, E. and Gutmann, S.},
	journal={Physical Review A},
	volume={58},
	pages={915--928},
	year={1998}
}

@article{aharonov1993,
	title={Quantum random walks},
	author={Aharonov, Y. and Davidovich, L. and Zagury, N.},
	journal={Physical Review A},
	volume={48},
	pages={1687--1690},
	year={1993}
}

@inproceedings{ambainis2001,
	title={One-dimensional quantum walks},
	author={Ambainis, A. and Bach, E. and Nayak, A. and Vishwanath, A. and Watrous, J.},
	booktitle={Proceedings of the thirty-third annual ACM symposium on Theory of computing, Hersonissos},
	year={2001}
}

@inproceedings{aharonov2001,
	title={Quantum walks on graphs},
	author={Aharonov, D. and Ambainis, A. and Kempe, J. and Vazirani, U.},
	booktitle={Proceedings of the thirty-third annual ACM symposium on Theory of computing, Hersonissos},
	year={2001}
}

@article{childs2009,
	title={Universal Computation by Quantum Walk},
	author={Childs, A. M.},
	journal={Physical Review Letters},
	volume={102},
	pages={180501},
	year={2009}
}

@article{asaka2021,
	title={Quantum random access memory via quantum walk},
	author={Asaka, R. and Sakai, K. and Yahagi, R.},
	journal={Quantum Science and Technology},
	volume={6},
	pages={035004},
	year={2021}
}

@article{asaka2023_1,
	title={Two-level quantum walkers on directed graphs. I. Universal quantum computing},
	author={Asaka, R. and Sakai, K. and Yahagi, R.},
	journal={Physical Review A},
	volume={107},
	pages={022415},
	year={2023}
}

@article{asaka2023_2,
	title={Two-level quantum walkers on directed graphs. II. Application to quantum random access memory},
	author={Asaka, R. and Sakai, K. and Yahagi, R.},
	journal={Physical Review A},
	volume={107},
	pages={022416},
	year={2023}
}

@article{schmidt2003,
	title={Realization of the Cirac--Zoller controlled-NOT quantum gate},
	author={Schmidt-Kaler, F. and H{\"a}ffner, H. and Riebe, M. and Gulde, S. and Lancaster, G. P. T. and Deuschle, T. and Becher, C. and Roos, C. F. and Eschner, J. and Blatt, R.},
	journal={Nature},
	volume={422},
	pages={408--411},
	year={2003}
}

@article{leibfried2003,
	title={Experimental demonstration of a robust, high-fidelity geometric two ion-qubit phase gate},
	author={Leibfried, D. and DeMarco, B. and Meyer, V. and Lucas, D. and Barrett, M. and Britton, J. and Itano, W. M. and Jelenkovi{\'c}, B. and Langer, C. and Rosenband, T. and Wineland, D. J.},
	journal={Nature},
	volume={422},
	pages={412--415},
	year={2003}
}

@article{kane1998,
	title={A silicon-based nuclear spin quantum computer},
	author={Kane, B. E.},
	journal={Nature},
	volume={393},
	pages={133--137},
	year={1998}
}

@article{pashkin2003,
	title={Quantum oscillations in two coupled charge qubits},
	author={Pashkin, Y. A. and Yamamoto, T. and Astafiev, O. and Nakamura, Y. and Averin, D. V. and Tsai, J. S.},
	journal={Nature},
	volume={421},
	pages={823--826},
	year={2003}
}

@article{obrien2003,
	title={Demonstration of an all-optical quantum controlled-NOT gate},
	author={O'Brien, J. L. and Pryde, G. J. and White, A. G. and Ralph, T. C. and Branning, D.},
	journal={Nature},
	volume={426},
	pages={264--267},
	year={2003}
}

@article{petrosyan2005,
	title={Towards deterministic optical quantum computation with coherently driven atomic ensembles},
	author={Petrosyan, D.},
	journal={Journal of Optics B: Quantum and Semiclassical Optics},
	volume={7},
	pages={S141},
	year={2005}
}

@article{andre2002,
	title={Manipulating Light Pulses via Dynamically Controlled Photonic Band gap},
	author={Andr{\'e}, A. and Lukin, M. D.},
	journal={Physical Review Letters},
	volume={89},
	pages={143602},
	year={2002}
}

@article{gasparoni2004,
	title={Realization of a Photonic Controlled-NOT Gate Sufficient for Quantum Computation},
	author={Gasparoni, S. and Pan, J.-W. and Walther, P. and Rudolph, T. and Zeilinger, A.},
	journal={Physical Review Letters},
	volume={93},
	pages={020504},
	year={2004}
}

@article{agarwal2005,
	title={Quantum random walk of the field in an externally driven cavity},
	author={Agarwal, G. S. and Pathak, P. K.},
	journal={Phys. Rev. A},
	volume={72},
	number={3},
	pages={033815},
	year={2005}
}

@article{cote2006,
	title={Quantum random walk with Rydberg atoms in an optical lattice},
	author={C{\^o}t{\'e}, R. and Russell, A. and Eyler, E. E. and Gould, P. L.},
	journal={New Journal of Physics},
	volume={8},
	pages={156},
	year={2006}
}

@article{jaksch2000,
	title={Fast Quantum Gates for Neutral Atoms},
	author={Jaksch, D. and Cirac, J. I. and Zoller, P. and Rolston, S. L. and C{\^o}t{\'e}, R. and Lukin, M. D.},
	journal={Physical Review Letters},
	volume={85},
	pages={2208--2211},
	year={2000}
}

@book{gallaghar1994,
	title={Rydberg Atoms},
	author={Gallaghar, T. F.},
	publisher={Cambridge University Press},
	year={1994}
}

@article{moller2008,
	title={Quantum Gates and Multiparticle Entanglement by Rydberg Excitation Blockade and Adiabatic Passage},
	author={M{\o}ller, D. and Madsen, L. B. and M{\o}lmer, K.},
	journal={Physical Review Letters},
	volume={100},
	pages={170504},
	year={2008}
}

@article{friedler2005,
	title={Deterministic quantum logic with photons via optically induced photonic band gaps},
	author={Friedler, I. and Kurizki, G. and Petrosyan, D.},
	journal={Physical Review A},
	volume={71},
	pages={023803},
	year={2005}
}

@article{shahmoon2011,
	title={Strongly interacting photons in hollow-core waveguides},
	author={Shahmoon, E. and Kurizki, G. and Fleischhauer, M. and Petrosyan, D.},
	journal={Physical Review A},
	volume={83},
	pages={033806},
	year={2011}
}

@article{petrosyan2012,
	title={Electromagnetically induced transparency and photon-photon interactions with Rydberg atoms},
	author={Petrosyan, D. and Fleischhauer, M.},
	journal={Journal of Physics: Conference Series},
	volume={350},
	pages={012001},
	year={2012}
}

@article{paredes2014,
	title={All-Optical Quantum Information Processing Using Rydberg Gates},
	author={Paredes-Barato, D. and Adams, C. S.},
	journal={Physical Review Letters},
	volume={112},
	pages={040501},
	year={2014}
}

@article{friedler2005_2,
	title={Long-range interactions and entanglement of slow single-photon pulses},
	author={Friedler, I. and Petrosyan, D. and Fleischhauer, M. and Kurizki, G.},
	journal={Phys. Rev. A},
	volume={72},
	number={4},
	pages={043803},
	year={2005}
}

@article{zeuner2018,
	title={Integrated-optics heralded controlled-NOT gate for polarization-encoded qubits},
	author={Zeuner, J. and Sharma, A. N. and Tillmann, M. and Heilmann, R. and Gr{\"a}fe, M. and Moqanaki, A. and Szameit, A. and Walther, P.},
	journal={npj Quantum Information},
	volume={4},
	pages={13},
	year={2018}
}

@article{xie2020,
	title={Design Rule of Mach-Zehnder Interferometer Sensors for Ultra-High Sensitivity},
	author={Xie, Y. and Zhang, M. and Dai, D.},
	journal={Sensors},
	volume={20},
	year={2020}
}

@article{zheng2023,
	title={Polarization Splitting at Visible Wavelengths with the Rutile TiO2 Ridge Waveguide},
	author={Zheng, X. and Ma, Y. and Zhao, C. and Xiang, B. and Yu, M. and Dai, Y. and Xu, F. and Lv, J. and Lu, F. and Zhou, C. and Ruan, S.},
	journal={Nanomaterials},
	volume={13},
	year={2023}
}

@article{cirac1997,
	title={Quantum State Transfer and Entanglement Distribution among Distant Nodes in a Quantum Network},
	author={Cirac, J. I. and Zoller, P. and Kimble, H. J. and Mabuchi, H.},
	journal={Phys. Rev. Lett.},
	volume={78},
	number={16},
	pages={3221--3224},
	year={1997}
}

@article{boozer2007,
	title={Reversible State Transfer between Light and a Single Trapped Atom},
	author={Boozer, A. D. and Boca, A. and Miller, R. and Northup, T. E. and Kimble, H. J.},
	journal={Phys. Rev. Lett.},
	volume={98},
	number={19},
	pages={193601},
	year={2007}
}
	
\end{document}